\pdfoutput=1

\documentclass[reprint,amsmath,amssymb,floatfix,longbibliography,aip,letterpaper,cha]{revtex4-1}

\usepackage{graphics,graphicx,float}
\usepackage{mathtools}		
\usepackage{comment}

\makeatletter
\preprintsty@sw{}{%
 \usepackage{titlesec}
 \titleformat{\section}{\normalfont\footnotesize\sffamily\bfseries\uppercase}%
 	{\thesection}{1em}{}%
 \titleformat{\subsubsection}{\normalfont\small\sffamily\slshape}{\thesubsubsection}{1em}{}%
 }%
\makeatother
  
\def\citeyear{\citep}
\def\autocite{\citep}

\usepackage[pdftex,
            pdfauthor={Steven A. Frank},
            pdftitle={d'Alembert's direct and inertial forces acting on populations: The Price equation and the fundamental theorem of natural selection},
            pdfsubject={d'Alembert's direct and inertial forces acting on populations: The Price equation and the fundamental theorem of natural selection}]{hyperref} 

\newcommand{\bdq}{\GD\mathbf{q}}
\newcommand{\bq}{\mathbf{q}}
\newcommand{\obq}{\mathbf{\dot{q}}}
\newcommand{\oq}{{\dot{q}}}
\newcommand{\ooq}{{\ddot{q}}}

\newcommand{\bdz}{\GD\mathbf{z}}
\newcommand{\bz}{\mathbf{z}}
\newcommand{\obz}{\mathbf{\dot{z}}}

\newcommand{\bmm}{\mathbf{m}}
\newcommand{\obm}{\mathbf{\dot{m}}}

\newcommand{\op}{{\dot{p}}}
\newcommand{\bp}{\mathbf{p}}
\newcommand{\obp}{\mathbf{\dot{p}}}
\newcommand{\oobp}{\mathbf{\ddot{p}}}

\newcommand{\zbar}{\bar{z}}
\newcommand{\mbar}{\bar{m}}

\newcommand{\ozbar}{\dot{\bar{z}}}

\newcommand{\ombar}{\dot{\bar{m}}}

\newcommand{\olog}[1]{\dot{\log #1}}

\newcommand{\orr}{{\dot{r}}}
\newcommand{\br}{\mathbf{r}}
\newcommand{\obr}{\mathbf{\dot{r}}}
\newcommand{\oobr}{\mathbf{\ddot{r}}}

\newcommand{\bF}{\mathbf{F}}
\newcommand{\bI}{\mathbf{I}}
\newcommand{\tF}{\tilde{\mathbf{F}}}
\newcommand{\tA}{\tilde{\mathbf{A}}}
\newcommand{\tI}{\tilde{\mathbf{I}}}

\newcommand{\norm}[1]{\left|#1\right|}
\newcommand{\sbq}{\sqrt{\bq}}
\newcommand{\sqi}{\sqrt{q_i}}

\newcommand{\HH}{\mathcal{H}}
\newcommand{\oHH}{\dot{\mathcal{H}}}

\newcommand{\Gbzm}{\Gb_{zm}}

\newcommand{\Gbzmd}{\Gb_{\dot{z}\dot{m}}}
\newcommand{\Geb}{\boldsymbol{\Ge}}
\newcommand{\Ggb}{\boldsymbol{\Gg}}

\newcommand{\bg}{\mathbf{g}}
\newcommand{\Gdb}{\boldsymbol{\Gd}}

\newcommand*{\Gb}{\beta}
\newcommand*{\Gd}{\delta}
\newcommand*{\GD}{\Delta}
\newcommand*{\Ge}{\epsilon}
\newcommand*{\Gg}{\gamma}

\newcommand*{\Gm}{\mu}

\newcommand*{\Gt}{\tau}

\newcommand*{\dd}{\textrm{d}}

\newcommand*{\Eq}[1]{eqn~\ref{eq:#1}}
\newcommand*{\Eqq}[1]{eqns~\ref{eq:#1}}

\newcommand*{\prt}{\partial}

\newcount\BoxNum \BoxNum 1\relax
\makeatletter
\newcommand*{\boxlabel}[1]{%
  \protected@write \@auxout {}{\string \newlabel {box:#1}{{\the\BoxNum}}{}}%
  \advance\BoxNum 1\relax}
\makeatother

\usepackage{titlesec}

\renewcommand{\thesection}{\arabic{section}}
\renewcommand{\thesubsection}{\thesection.\arabic{subsection}}
\renewcommand{\thesubsubsection}{\thesubsection.\arabic{subsubsection}}

\begin{document}

\title{d'Alembert's direct and inertial forces acting on populations: The Price equation and the fundamental theorem of natural selection}

\author{Steven A.\ Frank}
\affiliation{Department of Ecology and Evolutionary Biology, University of California, Irvine, CA 92697--2525  USA}

\begin{abstract}

I develop a framework for interpreting the forces that act on any population described by frequencies. The conservation of total frequency, or total probability, shapes the characteristics of force.  I begin with Fisher's fundamental theorem of natural selection. That theorem partitions the total evolutionary change of a population into two components. The first component is the partial change caused by the direct force of natural selection, holding constant all aspects of the environment. The second component is the partial change caused by the changing environment. I demonstrate that Fisher's partition of total change into the direct force of selection and the forces from the changing environmental frame of reference is identical to d'Alembert's principle of mechanics, which separates the work done by the direct forces from the work done by the inertial forces associated with the changing frame of reference. In d'Alembert's principle, there exist inertial forces from a change in the frame of reference that exactly balance the direct forces. I show that the conservation of total probability strongly shapes the form of the balance between the direct and inertial forces. I then use the strong results for conserved probability to obtain general results for the change in any system quantity, such as biological fitness or energy. Those general results derive from simple coordinate changes between frequencies and system quantities. Ultimately, d'Alembert's separation of direct and inertial forces provides deep conceptual insight into the interpretation of forces and the unification of disparate fields of study\footnote{\href{http://dx.doi.org/10.3390/e17107087}{doi:10.3390/e17107087} in \textit{Entropy}}\footnote{web: \href{http://stevefrank.org}{http://stevefrank.org}}.  

\bigskip


\end{abstract}

\maketitle



\section{Introduction}

The fundamental theorem of natural selection divides total evolutionary change into two components \autocite{fisher58the-genetical}. The first component is the partial change caused by the direct force of natural selection. The second component is the partial change caused by all other forces.

The theorem states that the change in fitness caused by the direct force of natural selection equals the genetic variance in fitness. We can interpret ``genetic variance'' to mean the component of variance associated with things that are transmitted through time. Natural selection is the force that changes the frequencies of those transmissible things. 

Fisher wrote clearly about the distinction between the direct force of natural selection and the other evolutionary forces \autocite{fisher58the-genetical,frank92fishers}. Yet much confusion followed in the history of the subject. Essentially all commentators considered only the total evolutionary change, rather than Fisher's split into two partial components. 

A correct interpretation of Fisher's partial components eventually developed, starting with Price \autocite{price72fishers} and Ewens \autocite{ewens89an-interpretation}. However, both of those authors concluded that Fisher's split of total change into components provided little value.

In this article, I show that Fisher's split of evolutionary change is equivalent to d'Alembert's split of the general causes of dynamics into direct and inertial forces. d'Alembert's principle is the foundation for essentially all of the key results of theoretical physics, starting with Newton's laws and leading to the subsequent generalizations via Lagrangian and Hamiltonian mechanics.

Lanczos \autocite{lanczos86the-variational}, in his great synthesis of the variational principles of mechanics, elevates d'Alembert's principle to the key insight that ties together the whole subject. To Lanczos, the tremendous value of d'Alembert's principle follows from the fact that it ``focuses attention on the forces, not on the moving body $\ldots$'' In the same way, Fisher's goal was to isolate and interpret the force of natural selection, rather than to emphasize the dynamics of total change. 

The study and interpretation of force requires separating the action of a force from the frame of reference. A force affects change, and the measurement and interpretation of that change depends on the changing frame of reference of the system. To understand the force as distinct from the frame of reference, force and frame of reference must be separated.

That separation between force and frame of reference is exactly what Fisher did and was exactly how he discussed his analysis. I argue here that connecting Fisher's theorem to d'Alembert's principle will help to clarify the separation of direct force and frame of reference. 

In Fisher's analysis, he was vague about the mathematical form of the changes associated with the frame of reference. Here, by using the Price equation, I make explicit the connections between Fisher's theorem and d'Alembert's principle.

My argument follows three steps. First, I derive the general form of the Price equation. Second, I connect the Price equation to d'Alembert's principle. Third, I discuss the fundamental theorem of natural selection in the context of d'Alembert's separation of the direct forces and the inertial forces associated with the changing frame of reference. By d'Alembert's separation, we obtain a partition of total evolutionary change in fitness into the change by the direct force of natural selection and the change by the inertial forces of the changing environmental frame of reference. 

The analysis is much more general and powerful than a theorem limited to natural selection. Instead, we find a broad analysis of the dynamics of any population or aggregation that can be characterized by frequencies. The conservation of total frequency, or total probability, establishes a symmetry that defines many of the characteristics of aggregate dynamics. Those characteristics of aggregate dynamics apply to natural selection, to many problems in mechanics, and to any analysis of the changes in probability distributions.  

\section{The Price equation}

The Price equation \autocite{price72extension,frank12naturalb} describes the change in an average value obtained over some aggregation or population. Each component of the population has a weighting, $q$, and a value, $z$. Begin with a discrete analog of the chain rule for differentiation of a product
\begin{align*}
  \GD(qz) &= (q+\GD q)(z+\GD z)-qz\\
          &= (\GD q)z+(q+\GD q)\GD z\\
          &= (\GD q)z+q'\GD z,
\end{align*}
in which $q'=q+\GD q$ and $z'=z+\GD z$. The same chain rule can be applied to vectors. By using dot product notation, we obtain an abstract form of the Price equation \autocite{frank12naturalb,frank12naturalc,frank13natural}
\begin{equation}\label{eq:priceGD}
  \GD(\bq\cdot\bz) = \bdq\cdot\bz + \bq'\cdot\bdz,
\end{equation}
in which a dot product is understood in the usual way as $\bq\cdot\bz=\sum q_iz_i$.

This equation can be interpreted in various ways. For our purposes, we can take $q_i$ to be the frequency associated with a subset, $i$, of the initial population, such that the total frequency is $\sum q_i=1$. Thus, $\zbar=\sum q_iz_i$ is the average of $z$, in which $z_i$ is a function that maps $i$ to some value. Similarly, we have a second population, with frequencies $q_i'$ and values $z_i'$, in which $\sum q_i'=1$. 

One can use various rules for the relations between $q_i$ and $q_i'$ and between $z_i$ and $z_i'$, allowing a wide variety of different perspectives on the transformations that relate the two populations \autocite{frank12naturalb}. For our purposes, we can operate abstractly and not worry about the particular rules. Our only restriction is that we can map the index $i$ between the two populations.

\section{Fitness as a change in frequency}

The function $z_i$ can map subset $i$ to any value. When studying frequency changes, let us rename the variable as $m\equiv z$, and choose 
\begin{equation*}
  m_i = \log\frac{q_i'}{q_i}
\end{equation*}
to describe the ratio of frequencies between the two populations associated with $i$. We can think of $m_i$ as a growth rate, or as a kind of force that moves the system from $q_i$ to $q_i'$. In particular, the above expression is equivalent to exponential growth driven by $m_i$ as
\begin{equation*}
  q_i' = q_i e^{m_i}.
\end{equation*}
We may call $m_i$ fitness, because it expresses the relative growth of the weighting associated with $i$. The term $m_i$ is, in effect, a growth rate relative to an unspecified underlying scale of change. We can take $m_i$ as a given force of growth and derive $q_i'$, or we can take the outcome $q_i'$ as given, and derive the effective force, $m_i$, that is consistent with the outcome. 

If we thought of $i$ as a particular individual or a particular type, then $m_i$ would express the growth rate associated with that individual or type between the two populations. However, the equations allow us simply to make the definition that relates $q_i$ to $q_i'$, and not restrict ourselves to a particular interpretation of what $i$ means in those terms.

I confine my analysis to small differences, $\GD q_i\rightarrow \dd q_i\equiv\oq_i$, in which $\oq_i=q_i'-q_i$ is small. For small differences we have (see Methods for assumptions)
\begin{equation*}
  m_i = \frac{\oq_i}{q_i}.
\end{equation*}
Using this definition and the substitution $m_i\equiv z_i$ in the Price equation \Eq{priceGD} from the prior section, we obtain a general expression for the total change in fitness as
\begin{equation*}
  \ombar = \obq\cdot\bmm + \bq\cdot\obm,
\end{equation*}
in which we ignore the second order term $\obq\cdot\obm$ in this description of small changes, with $\GD\bz\rightarrow\dd\bz\equiv\obm$.

\section{Conservation of total probability, entropy momentum, and Fisher information}

With the definition of fitness as a growth rate, $m_i=\oq_i/q_i$, average fitness is
\begin{equation*}
  \mbar=\bq\cdot\bmm=\sum \oq_i = 0.
\end{equation*}
This equation expresses the conservation of total probability or total frequency. It follows that the change in average fitness, $\ombar$, must also be zero
\begin{equation}\label{eq:ombar}
  \ombar = \obq\cdot\bmm + \bq\cdot\obm=0.
\end{equation}
The term $\obq\cdot\bmm$ has a wide variety of interpretations related to information theory and classical mechanics. For example, this term expresses entropy momentum or Fisher information \autocite{frieden04science,amari00methods}, as 
\begin{equation*}
  \obq\cdot\bmm = \sum \oq_i\,\olog{q_i} = \sum \frac{\oq_i^2}{q_i}.
\end{equation*}
The term $m_i=\olog{q_i} = \log{q_i'/q_i}$ is the change in entropy in each dimension, $i$, describing an entropy velocity or nondimensional entropy momentum relative to an unspecified underlying scale of change. Thus, $\obq\cdot\bmm$ may be interpreted as the gain in entropy momentum, which must be balanced by the loss of entropy momentum in the second term, $\bq\cdot\obm$, to achieve overall conservation, $\ombar=0$. 

Note that I have used $-\log q_i$ as the entropy in each dimension, consistent with the information theory concept of self-information or surprise as $-\log q_i$. That definition leads to system entropy as the expectation over the different dimensions, $-\sum q_i\log q_i$. Some people prefer to define the entropy in each dimension as $-q_i\log q_i$, and system entropy as the sum over each dimension, in which case my usage of entropy or information momentum does not make sense. 

The term $\sum\oq_i^2/q_i$ is widely used as the Fisher information metric, particularly in the study of information geometry \autocite{amari00methods}. Thus, the first term in $\ombar=0$ is the gain in Fisher information, and the second term is an exact balancing loss in Fisher information. The balance leads to an overall conservation of Fisher information, as emphasized by Frieden \autocite{frieden04science}.

We have transcended our original formulation of biological fitness in these descriptions of probability, information, and entropy. The expressions here apply to any problem that can be expressed in terms of changing frequencies in populations or aggregates, subject to the conservation of total frequency.

\section{d'Alembert's principle}

We may write d'Alembert's principle \autocite{lanczos86the-variational} as
\begin{equation*}
  \left(\bF + \bI\right)\obq=0.
\end{equation*}
Here, all terms are vectors, and the implicit dot product with $\obq$ distributes over the parentheses. The vector $\bq$ locates the system, and $\obq$ is a virtual displacement of the system from its current location to a nearby location. A virtual displacement is like an imaginary displacement, in which the system is held fixed in its current state, and then one moves its location without changing anything else. All forces and the frame of reference for measurement are held constant \autocite{lanczos86the-variational}. 

A virtual displacement must be consistent with all forces of constraint. In our case, the primary force of constraint on a virtual displacement, $\obq$, is that the sum of the frequencies is one. Thus, $\sum\oq_i=0$ expresses the force of constraint set by the conservation of total frequency or probability. Because a virtual displacement must be consistent with the forces of constraint, we need only analyze those forces that are in addition to the forces of constraint. In particular, we need to track the direct forces, $\bF$, and inertial forces, $\bI$.

The term $\bF$ is the vector of direct forces acting on the system, and the term $\bI$ is the vector of inertial forces that balance the direct forces to achieve no net change. d'Alembert's principle can be thought of as a generalization of Newton's second law of motion \autocite{lanczos86the-variational}, in which $\tF=\Gm \tA$ is read as the total force, $\tF$, equals mass, $\Gm$, times total acceleration, $\tA$. Total force and total acceleration must include forces of constraint. If we write total inertial force as $\tI=-\Gm\tA$, then Newton's law is $\tF+\tI=0$. 

When we study an actual system, we are usually interested in how the direct, or applied, forces influence dynamics. To do that, we need to separate the direct forces from the constraining forces. For example, in studying the frequency dynamics and evolutionary change caused by natural selection, we usually wish to analyze the direct force of growth rate, or fitness, separately from the force of constraint imposed by the conservation of total probability. 

In d'Alembert's formulation, the direct and inertial forces typically do not sum to zero, $\bF+\bI\ne0$, because those terms do not include the constraining forces. Instead, in d'Alembert's expression $\left(\bF + \bI\right)\obq=0$, the term $\obq\cdot\bF$ combines the direct and constraining forces, and the term $\obq\cdot\bI$ combines all inertial forces, including any forces of constraint. Newton's law is a special case of the more general principle of d'Alembert \autocite{lanczos86the-variational}.

\section{Interpretation of d'Alembert's principle}

Here is a simple intuitive description of d'Alembert's principle \autocite{wikipedia15fictitious}. You are sitting in a car at rest, and the car suddenly accelerates. You feel thrown back into the seat. But, even as the car gains speed, you effectively do not move in relation to the frame of reference of the car: your velocity relative to the car remains zero. That net zero velocity can be thought of as the balance between the direct force of the seat pushing on you and the inertial force sending you back as the car accelerates forward.

As long as your frame of reference moves with you, then your net motion in your frame of reference is zero. Put another way, there is always a changing frame of reference that zeroes net change by balancing the work of direct forces on a system against the work of a balancing inertial force. Although the system is a dynamic expression of changing components, it also has an overall static, equilibrium quality that aids analysis. As Lanczos \autocite{lanczos86the-variational} emphasizes, d'Alembert's principle ``focuses attention on the forces, not on the moving body $\ldots$''

\section{d'Alembert and the conservation of total probability}

This section transforms the conservation of total probability expressed by \Eq{ombar} into a form of d'Alembert's principle. We first note that (see Methods for $\olog{\bmm}$ notation)
\begin{equation*}
  \bq\cdot\obm = \left(\frac{\bq}{\obq}\odot\obm\right)\obq 
    = \left(\frac{\obm}{\bmm}\right)\obq = \olog{\bmm}\cdot\obq.
\end{equation*}
The symbol ``$\odot$'' denotes element-wise multiplication of vectors, the ratio denotes element-wise division, and dot products distribute over parentheses. With this expression, we can rewrite our general result in \Eq{ombar} for the conservation of total probability, or the change in fitness, in the general form of d'Alembert, $\left(\bF + \bI\right)\obq=0$, as
\begin{equation}\label{eq:dalemb}
  \left(\bmm + \olog{\bmm}\right)\obq=0.
\end{equation}
We equate this expression with d'Alembert by interpreting $\bmm\equiv\bF$ as the force of growth, or fitness, or, more generally, the direct forces acting on frequency change. We interpret $\olog{\bmm}\equiv\bI$ as the inertial forces, which typically are described in terms of acceleration with respect to the frame of reference. 

\section{Direct and inertial forces}

The expression in \Eq{dalemb} describes d'Alembert's principle for systems that follow conservation of total probability.  This section considers how we should interpret $\left(\bF+\bI\right)\obq=0$ for the direct and inertial forces in terms of Newtonian concepts of force and acceleration. 

The dot product expression in \Eq{dalemb} can be written as a sum over the individual dimensions of the system
\begin{equation*}
  \left(\bmm + \olog{\bmm}\right)\obq = \sum\left(m_i+\olog{m_i}\right)\oq_i.
\end{equation*}
The first term on each side, $\obq\cdot\bmm\equiv\obq\cdot\bF$, is the virtual displacement times the direct force. We may call this term the virtual work of the direct forces, because physical work is displacement times force. We can write this component of virtual work solely in terms of frequencies from our prior definition of $m_i=\oq_i/q_i$. 

The second term on each side, $\obq\cdot\olog{\bmm}\equiv\obq\cdot\bI$, is the virtual work of the inertial forces. To interpret the inertial forces with respect to acceleration, it is useful to express $\olog{\bmm}$ as
\begin{equation}\label{eq:logm}
  \olog{m_i} = \frac{\ooq_i}{\oq_i} - \frac{\oq_i}{q_i}.
\end{equation}
The term $\ooq_i$ is the second order infinitesimal change, or acceleration. Thus, $\bI\equiv\olog{\bmm}$ expresses how the changing frame of reference, arising from changed frequencies, leads to inertial forces that are accelerations. 

We can now write d'Alembert's principle under the conservation of total probability solely in terms of the probabilities, or frequencies, as
\begin{equation}\label{eq:dsum}
  \left(\bmm + \olog{\bmm}\right)\obq 
   = \sum\left(\frac{\oq_i}{q_i}+ \frac{\ooq_i}{\oq_i} - \frac{\oq_i}{q_i}\right)\oq_i=0.
\end{equation}
Distributing the virtual displacement, $\oq_i$, across the parentheses in the sum and splitting the sum into direct and inertial components yields
\begin{equation}\label{eq:dsum2}
  \sum\frac{\oq_i^2}{q_i}+ \sum\left(\ooq_i - \frac{\oq_i^2}{q_i}\right)=\sum \ooq_i=0.
\end{equation}
The sum of $\ooq_i$ is zero because $\sum\oq_i=0$ by conservation of total probability, and thus the accelerations, $\ooq_i$, also sum to zero. However, in a particular dimension, there may be an imbalance between direct and inertial force, $\ooq_i$. That imbalance arises because the force of constraint on total probability differs across dimensions.  

\section{Unitary coordinates and path lengths}

From \Eqq{dsum} and \ref{eq:dsum2}, we may express d'Alembert's balance between the total direct and inertial components as
\begin{equation}\label{eq:dsum3}
  \left(\bmm + \olog{\bmm}\right)\obq
   = \sum\frac{\oq_i^2}{q_i} - \sum\frac{\oq_i^2}{q_i}=0.
\end{equation}
The $\sum\oq_i^2/q_i$ terms can be understood as distances by considering the curvature caused by the constraining force of the conservation of total probability. To get a proper sense of distance in that curved geometric space, we need to change the coordinates. 

Let the new coordinates be $\br=\sqrt{\bq}$. Then the total Euclidean length of the vector $\br$ is the square root of the sum of squares in each dimension, which is 
\begin{equation*}
  \norm{\br}=\sqrt{\sum r_i^2}=\sqrt{\sum q_i}=1.
\end{equation*}
Vector lengths in the new coordinates are always one, which provides a pure expression of the conservation of total probability. In general, the $\bq$ may be arbitrary weightings, such that $\sum q_i$ is conserved, and thus $\sum\oq_i=0$. Here, I focus on conserved probability, in which the $q_i$ are positive and sum to one.

The path lengths of motion take on simple interpretations in terms of distance in the unitary coordinates. The transformed coordinates yield
\begin{equation*}
  \sum\frac{\oq_i^2}{q_i} = 4\sum \orr_i^2,
\end{equation*}
which shows the simple Euclidean interpretation of squared distance in the $\br$ coordinates as a sum of squared differences. This expression of distance is also equivalent to the Fisher information metric \autocite{frieden04science,amari00methods}. However, geometry is perhaps more fundamental than information, because the distance arises inevitably from curvature of paths caused by analyzing probability displacement subject to unitary conservation of total probability. 

\section{Geometry}

This section briefly reviews the geometry of frequency change dynamics that follow from two assumptions. The first assumption is that direct force, $m_i$, causes exponential growth
\begin{equation*}
  q_i'=q_ie^{m_i}.
\end{equation*}
This growth expression establishes a natural logarithmic scaling for comparing frequencies, because
\begin{equation*}
  m_i=\log\frac{q_i'}{q_i}.
\end{equation*}
When changes are small, $m_i=\olog{q_i}=\oq_i/q_i$. We could interpret those changes with respect to $\log q_i$ as entropy or information. But the geometry of force and growth may be a better way to think about the fundamental nature of these expressions.

The second assumption is that total frequency or probability is conserved, $\sum\oq_i=0$. That conservation imposes a constraint on paths of change. The constraint may be expressed by the geometry of the unitary coordinates, $\br=\sqrt{\bq}$, which yields a conserved length $\norm{\br}=1$. The path lengths for virtual displacements times direct or inertial forces are $\sum\oq_i^2/q_i = 4\sum \orr_i^2$. The essential geometry arising from growth and from conservation of total probability sets the form of the distances.

\section{Canonical coordinates and conservation in each dimension}

Hamiltonian expressions in canonical coordinates often provide the deepest insight into the symmetries of a system \autocite{landau76mechanics}. To obtain the Hamiltonian, the use of $\br=\sqrt{\bq}$ coordinates was a first step, because we can rewrite d'Alembert's principle in \Eq{dsum3} as
\begin{equation*}
  \frac{1}{4}\left(\bmm + \olog{\bmm}\right)\obq
   = \sum \orr_i^2 - \sum \orr_i^2=0.
\end{equation*}
However, the net balance only applies to the total system rather than separately in each dimension. If we can find the proper canonical coordinates, then the forces of constraint will appear independently in each dimension, and the balance of direct and inertial forces will also appear independently in each dimension.

In a Hamiltonian formulation, we assign two values to each component, usually considered as position and momentum \autocite{landau76mechanics}. In our nondimensional system, our primary factor is the conservation of total probability, which we express through the unitary coordinates $\br=\sbq$, such that the length of $\br$ is always one
\begin{equation*}
  \norm{\br}=\sbq\cdot\sbq=1.
\end{equation*}
If, for each point, we take $r_i=\sqi$ for position and $p_i=\sqi$ for momentum, then $\br\cdot\bp=1$, and the conserved Hamiltonian is
\begin{equation*}
  \HH = \obr\cdot\bp - \br\cdot\obp = 0.
\end{equation*}
This expression satisfies the requirements for Hamiltonian canonical coordinates of position and momentum, which are that $\prt\HH/\prt r_i=-\op_i$ and $\prt\HH/\prt p_i=\orr_i$.
The differential of the Hamiltonian often provides a useful expression
\begin{equation}\label{eq:hdot}
  \oHH=\oobr\cdot\bp - \br\cdot\oobp=0,
\end{equation}
which, in each separate dimension, is zero
\begin{equation}\label{eq:hi}
  \oHH_i= \ddot{r_i}p_i-r_i\ddot{p}_i=0,
\end{equation}
because $r_i=p_i=\sqi$, and 
\begin{equation*}
  \ddot{r_i}=\ddot{p_i}=\frac{1}{2\sqi}\left(\ddot{q_i}-\frac{\oq_i^2}{2q_i}\right),
\end{equation*}
thus we can write the Hamiltonian in each dimension as
\begin{equation*}
  4\oHH_i=\left(\frac{\oq_i^2}{q_i}-2\ddot{q_i}\right)
     -\left(\frac{\oq_i^2}{q_i}-2\ddot{q_i}\right)=0.
\end{equation*}
Here, the curvature from the force of constraint is divided into equal and opposite contributions in the direct and inertial force components, recovering a Newtonian $\tF_i-\Gm\tA_i=0$ perspective independently in each dimension.

We can rewrite \Eq{hdot} as a d'Alembert's principle expression
\begin{equation*}
  \oHH = \left(\bp\odot\olog{\,}\obr - \br\odot\olog{\,}\obp\right)\obr=0,
\end{equation*}
for virtual displacement $\obr$, direct force $\bF=-\bp\odot\olog{\,}\obr$, and inertial force $\bI=\br\odot\olog{\,}\obp$. The symbol ``$\odot$'' denotes element-wise multiplication of vectors, and dot products distribute over parentheses. Thus, $\oHH=\left(\bF+\bI\right)\obr=0$, with the Newtonian equality $\bF_i+\bI_i=0$ satisfied in each dimension.

\section{Coordinates for quantities correlated with force}

We can analyze any quantitative system property by transforming coordinates. We start with the general results for the conservation of total probability and information momentum, $\ombar=0$. We then obtain an expression for the change in the system quantity, $\ozbar$, by the change in coordinates $(\bmm,\obm)\mapsto (\bz,\obz)$, in which the different coordinates now have an arbitrary relation rather than the earlier equivalence. That change in coordinates generalizes the $\ombar$ form of the Price equation (\Eq{ombar}), to give the change in the average value of $z$ as
\begin{equation*}
  \ozbar=\obq\cdot\bz + \bq\cdot\obz.
\end{equation*}
The $z_i$ values are the averages of $z$ in each dimension, $i$. Because $z$ can be any quantity, calculated in any way, this equation gives the most general expression for $\ozbar$, the change in the average of $z$. One can think of $\zbar=\sum q_i z_i$ as a functional of the arbitrary function, $z$, that maps $i\mapsto z_i$. The only restriction on the expression for $\ozbar$ shown here is that changes be small. For large changes, the exact form of the Price equation in \Eq{priceGD} should be used. 

We can relate $\ombar$ to $\ozbar$ by writing the change in coordinates, $\bmm\mapsto\bz$ and $\obm\mapsto\obz$, as the regression equations
\begin{align*}
  \bz &= \Gbzm \bmm +\Geb\\
  \obz &= \Gbzmd \obm + \Ggb,
\end{align*}
in which the regression coefficients, $\Gb$, are obtained by minimizing the length of the ``error'' vector. To analyze the length of the error vector, we can use standard identities from the theory of least squares for regression \autocite{draper98applied}. 

In particular, the first regression equation follows from choosing $\Gbzm$ to minimize $\norm{\Geb_\bq}^2=\sum q_i\Ge_i^2$, in which $\Geb_\bq=\sqi\Ge_i$ denotes a $\sqrt{\bq}$ weighted vector. Choosing $\Gbzm$ to minimize the length of $\Geb_\bq$ leads to $\bmm_\bq\cdot\Geb_\bq=0$, because the minimum length of $\Geb_\bq$ occurs when that vector is orthogonal to $\bmm_\bq$. Note that $\oq_i=q_im_i$, thus
\begin{equation*}
  \obq\cdot\Geb=\sum q_im_i\Ge_i=\bmm_\bq\cdot\Geb_\bq=0.
\end{equation*}
In the equation for $\obz$, minimizing $\norm{\Ggb_\bq}^2$ sets $\Gbzmd$. We also have, by standard theory, $\bq\cdot\Ggb=0$.
 
Using these identities, 
\begin{align}
  \obq\cdot\bz &= \Gbzm\obq\cdot\bmm + \obq\cdot\Geb=\Gbzm\obq\cdot\bmm
  	\label{eq:qdotz}\\
  \bq\cdot\obz &= \Gbzmd\bq\cdot\obm + \bq\cdot\Ggb= \Gbzmd\bq\cdot\obm,
  	\label{eq:qzdot}
\end{align}
from which we obtain the change $\ozbar$ in terms of the original coordinates for $\ombar$ as
\begin{equation}\label{eq:ozbarReg}
  \ozbar = \Gbzm\obq\cdot\bmm + \Gbzmd\bq\cdot\obm = \left(\Gbzm-\Gbzmd\right)\obq\cdot\bmm,
\end{equation}
the right expression arising from the fact that $\obq\cdot\bmm + \bq\cdot\obm=0$. The total change, $\ozbar$, is split into the virtual work term, $\Gbzm\obq\cdot\bmm$, and the inertial force term, $\Gbzmd\bq\cdot\obm$. The regression coefficients rescale coordinates $(\bmm,\obm)\mapsto(\bz,\obz)$.

If $\zbar$ is a conserved quantity, or the system is at an equilibrium with respect to $\zbar$, then $\ozbar=0$. We can write a d'Alembert form
\begin{equation*}
  \ozbar=\left(\Gbzm\bmm-\Gbzmd\bmm\right)\obq=0
\end{equation*}
which, when $\obq\cdot\bmm\ne0$, implies $\Gbzm=\Gbzmd$, and the d'Alembert equality holds separately in each dimension. In this case, the dynamics of $\bz$ are influenced by both the conservation of probability and by additional constraints set by the conservation of $\zbar$. We may, of course, choose the changing reference frame, $\obz$, such that $\ozbar\ne0$, in which case the direct and inertial forces do not completely balance. 

\section{The fundamental theorem}

We may set $\Gbzmd=0$, either because the changing value of $\zbar$ is unaffected by the changing reference frame, or because the effects of the changing reference frame are ignored by assumption. We then have an expression for the partial change caused by the direct forces, holding constant the frame of reference
\begin{equation*}
  \ozbar_s=\obq\cdot\bz=\Gbzm\obq\cdot\bmm,
\end{equation*}
in which the $s$ subscript emphasizes that this is a partial change ascribed to the direct forces, or the forces of selection. This form includes, as special cases, Fisher's fundamental theorem of natural selection, the breeder's equation of genetics, and other common expressions for the change in populations caused by natural selection. 

Note that $\obq\cdot\bmm=V_m$, the variance of $\bmm$, because 
\begin{equation*}
  \obq\cdot\bmm=\sum\oq_i m_i=\sum q_i\left(\frac{\oq_i}{q_i}\right)m_i=\sum q_im_i^2,
\end{equation*}
which is the variance of $\bmm$, because $\mbar=0$.

If we take $\bz=\bmm$ in order to study the change in fitness caused by the direct forces, then $\ombar_s=V_m$, the change in mean fitness caused by selection, $\ombar_s$, is the variance in fitness, $V_m$. Fisher was interested in the transmissible change in $\mbar$ associated with genetic factors, $\bg$, thus he partitioned fitness as $\bmm=\bg+\Gdb$. Here, the genetic factors are partial regressions associated with particular genes, such that $\bg$ is chosen to maximize the amount of the total variance in fitness, $V_m$, associated with the transmissible genes \autocite{ewens89an-interpretation,frank13natural,frank97the-price,crow70an-introduction}. The $\Gdb$ terms are residuals in the regression, such that one gets the additive partition of total variance from classical regression theory as $V_m=V_g+V_\Gd$. 

The change in fitness caused by the direct forces can now be written as
\begin{equation*}
  \ombar_s=V_g+V_\Gd,
\end{equation*}
and thus the transmissible change in fitness caused by natural selection and associated with genetic factors is 
\begin{equation*}
  \ombar_{s|g}=V_g,
\end{equation*}
in which $V_g$ is the variance in the transmissible effects of the genetic factors on fitness, or the genetic variance in fitness. That partial change in fitness caused by direct forces and associated with transmissible factors is what Fisher emphasized in his fundamental theorem of natural selection. By defining the genetic factors, $\bg$, as the only direct forces of interest, the residual forces of selection, $\Gdb$, are added to the other inertial forces that define the changing frame of reference. 

In models of evolutionary change, Fisher chose to ascribe the direct force of change associated with $\bg$ to natural selection, and all other forces to the inertial frame that he called environmental causes. That d'Alembert interpretation of the split between direct and inertial forces provides a clear way in which to understand Fisher's fundamental theorem of natural selection. There is, of course, an arbitrary aspect to such a partition, because the split between direct and inertial forces depends entirely on how one chooses to define the frames of reference. For example, a change in how one defines the set of potentially transmissible factors, $\bg$, alters how one splits forces between direct and inertial components \autocite{frank97the-price}.

\section{Conclusions}

The fundamental equations for change are identical between many laws of physics and evolutionary change by natural selection. However, the different histories of those subjects and the long and confused debates in biology about Fisher's fundamental theorem have obscured the simple, common basis of the underlying theory.

I unified different theories by combining d'Alembert's conceptual frame with the abstract expressions of the Price equation. That combination led to a simple and very general basis for understanding populations or aggregations, in which one can interpret total frequency or total probability as a conserved quantity. By combining conservation of total frequency with a notion of change based on exponential growth, I showed the geometric and algebraic forms of change that arise from d'Alembert's partition of direct and inertial forces. I also provided an elegant Hamiltonian expression in canonical coordinates, which recovers the Newtonian balance of force and acceleration independently in each dimension for the corresponding direct and inertial forces of d'Alembert.

Finally, I showed that arbitrary system quantities, such as biological traits, or any total system quantity such as energy, can be interpreted through two steps. First, begin with the universal results that arise from conservation of total probability and the notion of change as exponential growth. Second, apply a simple coordinate transformation between frequency change and system quantities to obtain general expressions for the change in system quantities.

\section{Methods}

The assumption of small changes associated with the overdot notation does not imply that forces are weak. Instead, the scale of change is small, in the sense typically associated with continuous time derivatives in differential equations. However, I have avoided classical derivative notation and differential equations in order to retain the more general form of the abstract Price equation \autocite{frank12naturalb,frank12naturalc}.  

For example, in the definition $m_i=\oq_i/q_i$, the overdot notation can be interpreted as a small change in $q_i$, such that $\oq_i\equiv\dd q_i$. Fitness in biology is sometimes given as an absolute number or as a nondimensional change in frequency, consistent with $m_i$, and sometimes as a rate or Malthusian parameter, which might be given as
\begin{equation}
  M_i = \frac{m_i}{\dd\Gt}=\frac{\oq_i}{q_i\dd\Gt}=\frac{\dd\log{q_i}}{\dd\Gt}.
\end{equation}
Here, $\dd\Gt$ is the underlying scale of change, which is typically a small change in time. However, we can take $\dd\Gt$ as an abstraction of the underlying scale of change, which may have any units or be nondimensional. If we take the units on $\Gt$ as the square of time, then we move toward traditional definitions of force or acceleration. Because $\dd\Gt$ is small, the quantities of rates, forces, or accelerations may be large. 

In the text, we are always looking at equivalences between left and right hand sides of equations. So we can always multiply or divide by various functions of $\dd\Gt$ interpreted with respect to arbitrary dimensions. The abstraction in the text is intentional, because the interdisciplinary connections between seemingly different subjects and results arise only when one focuses on the abstract structure of the key results. For example, the need for such abstraction arose elsewhere when studying the relation between Fisher's fundamental theorem and Fisher information \autocite{frank12naturalb,frank12naturalc,frank09natural}. 

The abstract structure shows the unity among a broad array of fundamental expressions in mechanics, in biology, in information theory and information geometry, and in many other kinds of problems that can be cast in variational form. 

I have made the assumption that the scale of change is small, and thus all quantities with overdots are small. In biology, that assumption is often associated with models of populations with overlapping generations described in continuous time differential equations \autocite{crow70an-introduction}. In mechanics, that assumption corresponds to the classical differential equation expressions in continuous time. 

The analysis of discrete changes that are not small, typically associated with discrete time models, remains an open problem. The exact Price expression in \Eq{priceGD} gives a hint at how to proceed when changes are not small. The connection to the continuous expressions of mechanics and d'Alembert might be achieved by careful use of differential geometry and construction of discrete changes as sums of small changes along continuous paths. But that analysis remains an open problem for the future. Some results based on the analysis of the exact, discrete Price equation may provide a point of departure \autocite{frank12naturalb,frank12naturalc}.

The $\olog{m_i}$ notation is interpreted as 
\begin{equation*}
  \olog{m_i}=\frac{\dd m_i}{m_i},
\end{equation*}
which is the change in the relative distance of $m_i$ from zero. This interpretation is consistent with the expression of $\olog{m_i}$ in terms of the changes in $q_i$ given in \Eq{logm}.

\begin{acknowledgments}
National Science Foundation grant DEB--1251035 supports my research.  I completed this work while on fellowship at the Wissenschaftskolleg zu Berlin.
\end{acknowledgments}

\bigskip
\bibliography{main}

\begin{thebibliography}{17}%
\makeatletter
\providecommand \@ifxundefined [1]{%
 \@ifx{#1\undefined}
}%
\providecommand \@ifnum [1]{%
 \ifnum #1\expandafter \@firstoftwo
 \else \expandafter \@secondoftwo
 \fi
}%
\providecommand \@ifx [1]{%
 \ifx #1\expandafter \@firstoftwo
 \else \expandafter \@secondoftwo
 \fi
}%
\providecommand \natexlab [1]{#1}%
\providecommand \enquote  [1]{``#1''}%
\providecommand \bibnamefont  [1]{#1}%
\providecommand \bibfnamefont [1]{#1}%
\providecommand \citenamefont [1]{#1}%
\providecommand \href@noop [0]{\@secondoftwo}%
\providecommand \href [0]{\begingroup \@sanitize@url \@href}%
\providecommand \@href[1]{\@@startlink{#1}\@@href}%
\providecommand \@@href[1]{\endgroup#1\@@endlink}%
\providecommand \@sanitize@url [0]{\catcode `\\12\catcode `\$12\catcode
  `\&12\catcode `\#12\catcode `\^12\catcode `\_12\catcode `\%12\relax}%
\providecommand \@@startlink[1]{}%
\providecommand \@@endlink[0]{}%
\providecommand \url  [0]{\begingroup\@sanitize@url \@url }%
\providecommand \@url [1]{\endgroup\@href {#1}{\urlprefix }}%
\providecommand \urlprefix  [0]{URL }%
\providecommand \Eprint [0]{\href }%
\providecommand \doibase [0]{http://dx.doi.org/}%
\providecommand \selectlanguage [0]{\@gobble}%
\providecommand \bibinfo  [0]{\@secondoftwo}%
\providecommand \bibfield  [0]{\@secondoftwo}%
\providecommand \translation [1]{[#1]}%
\providecommand \BibitemOpen [0]{}%
\providecommand \bibitemStop [0]{}%
\providecommand \bibitemNoStop [0]{.\EOS\space}%
\providecommand \EOS [0]{\spacefactor3000\relax}%
\providecommand \BibitemShut  [1]{\csname bibitem#1\endcsname}%
\let\auto@bib@innerbib\@empty
\bibitem [{\citenamefont {Fisher}(1958)}]{fisher58the-genetical}%
  \BibitemOpen
  \bibfield  {author} {\bibinfo {author} {\bibfnamefont {R.~A.}\ \bibnamefont
  {Fisher}},\ }\href@noop {} {\emph {\bibinfo {title} {The {G}enetical {T}heory
  of {N}atural {S}election}}},\ \bibinfo {edition} {2nd}\ ed.\ (\bibinfo
  {publisher} {Dover},\ \bibinfo {address} {New York},\ \bibinfo {year}
  {1958})\BibitemShut {NoStop}%
\bibitem [{\citenamefont {Frank}\ and\ \citenamefont
  {Slatkin}(1992)}]{frank92fishers}%
  \BibitemOpen
  \bibfield  {author} {\bibinfo {author} {\bibfnamefont {S.~A.}\ \bibnamefont
  {Frank}}\ and\ \bibinfo {author} {\bibfnamefont {M.}~\bibnamefont
  {Slatkin}},\ }\bibfield  {title} {\enquote {\bibinfo {title} {Fisher's
  fundamental theorem of natural selection},}\ }\href@noop {} {\bibfield
  {journal} {\bibinfo  {journal} {Trends in Ecology and Evolution}\ }\textbf
  {\bibinfo {volume} {7}},\ \bibinfo {pages} {92--95} (\bibinfo {year}
  {1992})}\BibitemShut {NoStop}%
\bibitem [{\citenamefont {Price}(1972{\natexlab{a}})}]{price72fishers}%
  \BibitemOpen
  \bibfield  {author} {\bibinfo {author} {\bibfnamefont {G.~R.}\ \bibnamefont
  {Price}},\ }\bibfield  {title} {\enquote {\bibinfo {title} {Fisher's
  `fundamental theorem' made clear},}\ }\href@noop {} {\bibfield  {journal}
  {\bibinfo  {journal} {Annals of Human Genetics}\ }\textbf {\bibinfo {volume}
  {36}},\ \bibinfo {pages} {129--140} (\bibinfo {year}
  {1972}{\natexlab{a}})}\BibitemShut {NoStop}%
\bibitem [{\citenamefont {Ewens}(1989)}]{ewens89an-interpretation}%
  \BibitemOpen
  \bibfield  {author} {\bibinfo {author} {\bibfnamefont {W.~J.}\ \bibnamefont
  {Ewens}},\ }\bibfield  {title} {\enquote {\bibinfo {title} {An interpretation
  and proof of the fundamental theorem of natural selection},}\ }\href@noop {}
  {\bibfield  {journal} {\bibinfo  {journal} {Theoretical Population Biology}\
  }\textbf {\bibinfo {volume} {36}},\ \bibinfo {pages} {167--180} (\bibinfo
  {year} {1989})}\BibitemShut {NoStop}%
\bibitem [{\citenamefont {Lanczos}(1986)}]{lanczos86the-variational}%
  \BibitemOpen
  \bibfield  {author} {\bibinfo {author} {\bibfnamefont {C.}~\bibnamefont
  {Lanczos}},\ }\href@noop {} {\emph {\bibinfo {title} {The Variational
  Principles of Mechanics}}},\ \bibinfo {edition} {4th}\ ed.\ (\bibinfo
  {publisher} {Dover Publications},\ \bibinfo {address} {New York},\ \bibinfo
  {year} {1986})\BibitemShut {NoStop}%
\bibitem [{\citenamefont {Price}(1972{\natexlab{b}})}]{price72extension}%
  \BibitemOpen
  \bibfield  {author} {\bibinfo {author} {\bibfnamefont {G.~R.}\ \bibnamefont
  {Price}},\ }\bibfield  {title} {\enquote {\bibinfo {title} {Extension of
  covariance selection mathematics},}\ }\href@noop {} {\bibfield  {journal}
  {\bibinfo  {journal} {Annals of Human Genetics}\ }\textbf {\bibinfo {volume}
  {35}},\ \bibinfo {pages} {485--490} (\bibinfo {year}
  {1972}{\natexlab{b}})}\BibitemShut {NoStop}%
\bibitem [{\citenamefont {Frank}(2012{\natexlab{a}})}]{frank12naturalb}%
  \BibitemOpen
  \bibfield  {author} {\bibinfo {author} {\bibfnamefont {S.~A.}\ \bibnamefont
  {Frank}},\ }\bibfield  {title} {\enquote {\bibinfo {title} {Natural
  selection. {IV}. {T}he {P}rice equation},}\ }\href@noop {} {\bibfield
  {journal} {\bibinfo  {journal} {Journal of Evolutionary Biology}\ }\textbf
  {\bibinfo {volume} {25}},\ \bibinfo {pages} {1002--1019} (\bibinfo {year}
  {2012}{\natexlab{a}})}\BibitemShut {NoStop}%
\bibitem [{\citenamefont {Frank}(2012{\natexlab{b}})}]{frank12naturalc}%
  \BibitemOpen
  \bibfield  {author} {\bibinfo {author} {\bibfnamefont {S.~A.}\ \bibnamefont
  {Frank}},\ }\bibfield  {title} {\enquote {\bibinfo {title} {Natural
  selection. {V}. {H}ow to read the fundamental equations of evolutionary
  change in terms of information theory},}\ }\href@noop {} {\bibfield
  {journal} {\bibinfo  {journal} {Journal of Evolutionary Biology}\ }\textbf
  {\bibinfo {volume} {25}},\ \bibinfo {pages} {2377--2396} (\bibinfo {year}
  {2012}{\natexlab{b}})}\BibitemShut {NoStop}%
\bibitem [{\citenamefont {Frank}(2013)}]{frank13natural}%
  \BibitemOpen
  \bibfield  {author} {\bibinfo {author} {\bibfnamefont {S.~A.}\ \bibnamefont
  {Frank}},\ }\bibfield  {title} {\enquote {\bibinfo {title} {Natural
  selection. {VI}. {P}artitioning the information in fitness and characters by
  path analysis},}\ }\href@noop {} {\bibfield  {journal} {\bibinfo  {journal}
  {Journal of Evolutionary Biology}\ }\textbf {\bibinfo {volume} {26}},\
  \bibinfo {pages} {457--471} (\bibinfo {year} {2013})}\BibitemShut {NoStop}%
\bibitem [{\citenamefont {Frieden}(2004)}]{frieden04science}%
  \BibitemOpen
  \bibfield  {author} {\bibinfo {author} {\bibfnamefont {B.~R.}\ \bibnamefont
  {Frieden}},\ }\href@noop {} {\emph {\bibinfo {title} {{Science from Fisher
  Information: A Unification}}}}\ (\bibinfo  {publisher} {Cambridge University
  Press},\ \bibinfo {address} {Cambridge, UK},\ \bibinfo {year}
  {2004})\BibitemShut {NoStop}%
\bibitem [{\citenamefont {Amari}\ and\ \citenamefont
  {Nagaoka}(2000)}]{amari00methods}%
  \BibitemOpen
  \bibfield  {author} {\bibinfo {author} {\bibfnamefont {S.}~\bibnamefont
  {Amari}}\ and\ \bibinfo {author} {\bibfnamefont {H.}~\bibnamefont
  {Nagaoka}},\ }\href@noop {} {\emph {\bibinfo {title} {{Methods of Information
  Geometry}}}}\ (\bibinfo  {publisher} {Oxford University Press},\ \bibinfo
  {address} {New York},\ \bibinfo {year} {2000})\BibitemShut {NoStop}%
\bibitem [{\citenamefont {Wikipedia}(2015)}]{wikipedia15fictitious}%
  \BibitemOpen
  \bibfield  {author} {\bibinfo {author} {\bibnamefont {Wikipedia}},\ }\href
  {http://en.wikipedia.org/w/index.php?title=Fictitious_force&oldid=659661243}
  {\enquote {\bibinfo {title} {Fictitious force --- {W}ikipedia{,} {T}he {F}ree
  {E}ncyclopedia},}\ } (\bibinfo {year} {2015}),\ \bibinfo {note} {[Online;
  accessed 22-May-2015]}\BibitemShut {NoStop}%
\bibitem [{\citenamefont {Landau}\ and\ \citenamefont
  {Lifshitz}(1976)}]{landau76mechanics}%
  \BibitemOpen
  \bibfield  {author} {\bibinfo {author} {\bibfnamefont {L.~D.}\ \bibnamefont
  {Landau}}\ and\ \bibinfo {author} {\bibfnamefont {E.~M.}\ \bibnamefont
  {Lifshitz}},\ }\href@noop {} {\emph {\bibinfo {title} {Mechanics}}},\
  \bibinfo {edition} {3rd}\ ed.,\ \bibinfo {series} {Course in Theoretical
  Physics}, Vol.~\bibinfo {volume} {1}\ (\bibinfo  {publisher}
  {Butterworth-Heinemann},\ \bibinfo {address} {London},\ \bibinfo {year}
  {1976})\BibitemShut {NoStop}%
\bibitem [{\citenamefont {Draper}\ and\ \citenamefont
  {Smith}(1998)}]{draper98applied}%
  \BibitemOpen
  \bibfield  {author} {\bibinfo {author} {\bibfnamefont {N.~R.}\ \bibnamefont
  {Draper}}\ and\ \bibinfo {author} {\bibfnamefont {H.}~\bibnamefont {Smith}},\
  }\href@noop {} {\emph {\bibinfo {title} {Applied Regression Analysis}}},\
  \bibinfo {edition} {3rd}\ ed.\ (\bibinfo  {publisher} {Wiley-Interscience},\
  \bibinfo {address} {Hoboken, NJ},\ \bibinfo {year} {1998})\BibitemShut
  {NoStop}%
\bibitem [{\citenamefont {Frank}(1997)}]{frank97the-price}%
  \BibitemOpen
  \bibfield  {author} {\bibinfo {author} {\bibfnamefont {S.~A.}\ \bibnamefont
  {Frank}},\ }\bibfield  {title} {\enquote {\bibinfo {title} {The {{P}}rice
  equation, {{F}}isher's fundamental theorem, kin selection, and causal
  analysis},}\ }\href@noop {} {\bibfield  {journal} {\bibinfo  {journal}
  {Evolution}\ }\textbf {\bibinfo {volume} {51}},\ \bibinfo {pages}
  {1712--1729} (\bibinfo {year} {1997})}\BibitemShut {NoStop}%
\bibitem [{\citenamefont {Crow}\ and\ \citenamefont
  {Kimura}(1970)}]{crow70an-introduction}%
  \BibitemOpen
  \bibfield  {author} {\bibinfo {author} {\bibfnamefont {J.~F.}\ \bibnamefont
  {Crow}}\ and\ \bibinfo {author} {\bibfnamefont {M.}~\bibnamefont {Kimura}},\
  }\href@noop {} {\emph {\bibinfo {title} {An {I}ntroduction to {P}opulation
  {G}enetics {T}heory}}}\ (\bibinfo  {publisher} {Burgess},\ \bibinfo {address}
  {Minneapolis, Minnesota},\ \bibinfo {year} {1970})\BibitemShut {NoStop}%
\bibitem [{\citenamefont {Frank}(2009)}]{frank09natural}%
  \BibitemOpen
  \bibfield  {author} {\bibinfo {author} {\bibfnamefont {S.~A.}\ \bibnamefont
  {Frank}},\ }\bibfield  {title} {\enquote {\bibinfo {title} {Natural selection
  maximizes {F}isher information},}\ }\href@noop {} {\bibfield  {journal}
  {\bibinfo  {journal} {Journal of Evolutionary Biology}\ }\textbf {\bibinfo
  {volume} {22}},\ \bibinfo {pages} {231--244} (\bibinfo {year}
  {2009})}\BibitemShut {NoStop}%
\end{thebibliography}%

\end{document}